30th International Symposium on Space Flight Dynamics (ISSFD) –

# Rosetta Earth Flyby Anomaly Revisited: Assessing Physical Contributions through Orbital Dynamics Analysis

Tobias Hoffmann(1), Frank Budnik(1)

(1) *Flight Dynamics Division, ESA/ESOC*
*Darmstadt, Germany*
*Emails: tobias.hoffmann@esa.int, frank.budnik@esa.int*

***Abstract* – Background: During the navigation of Rosetta's first Earth flyby in 2005, a small yet unexplained asymptotic velocity increase was detected in the departing leg, necessitating inserting an artificial prograde delta-V at perigee in the orbit determination. Similar instances of this phenomenon have been observed in several missions. However, subsequent Rosetta Earth flybys did not exhibit this effect. Despite numerous investigations of the – so called Earth flyby anomaly – over the last decades, the underlying cause remains unresolved.**

**Idea: To address the problem, we perform an Orbit Determination using ESOC's new system for Deep Space orbit determination, utilising the original radiometric tracking data and spacecraft model. This approach enables the incorporation of additional physical effects in the flight trajectory, with the aim of identifying the root cause of the anomaly.**

**Purpose: Achieving high-accuracy navigation during gravity-assist manoeuvres, while refining physical models and addressing unresolved dynamical effects, integrates operational and scientific objectives that are of relevance for current and future missions.**

**Results: We successfully reproduced the anomaly using the new orbit determination software. Subsequent analysis ruled out explanations involving Earth ephemeris errors, gravitational field modelling, relativistic contributions, and propagator configurations. Additional dynamical variations, such as the solar radiation pressure or albedo radiation, were evaluated but found insufficient to account for the observed discrepancy.**

**Outlook: During investigations of Earth's albedo effect, a possible explanation for an artificial velocity increase was identified. Albedo-induced illumination on thermally sensitive spacecraft surfaces could have triggered outgassing, consistent in direction and magnitude with documented outgassing events during Rosetta's cruise phase. A detailed attitude analysis further strengthens this theory and provides conclusive results, also in comparison to Rosetta's other Earth flybys.**

## I. INTRODUCTION

Planetary flybys are regularly used in interplanetary trajectory design, exploiting the gravity of planetary bodies to change the heliocentric velocity of a spacecraft without using propellant, thus also being called "gravity-assist manoeuvres". They efficiently exploit the energy and momentum transfer of both objects: In the planetary frame the spacecraft is on a hyperbolic orbit, changing its direction because of gravitational pull, with the asymptotic velocity $V_\infty$ remaining the same in the in- and outgoing leg. But due to the relative speed of the planetary object, the spacecraft's velocity vector effectively changes with respect to the Sun, creating the slingshot effect. Surprisingly, during several Earth flybys, an unexplained asymptotic velocity increase $\Delta V_\infty$ in the order of a few mm/s could be clearly measured.

The anomalous velocity increase at Earth flybys was first detected for the NASA spacecraft Galileo (1990) and NEAR (1998) [1], and was later observed again in 2005 by ESA's Rosetta mission [2], but its origin remains unexplained for more than 30 years. Since then, several more Earth gravity-assist manoeuvres have been completed by Rosetta and other spacecraft (e.g. [3–5]], yet none could significantly measure again this anomalous behaviour, only the recent JUICE combined Lunar-Earth Gravity-Assist (LEGA) in 2024 suggests a small measurable prograde velocity increase of around 0.1 mm/s at Earth's closest approach [6]. A summary of spacecraft Earth flybys and their measured ΔV can be found in Table 1.

There have been various attempts to solve the anomaly. Empirically derived formulas have been proposed by Anderson et al. [3] and later improved versions based on statistical analysis [4]. However, both failed to accurately predict upcoming flybys and explain null results. Neither of these provides a physical explanation for the anomaly. On the other hand, further analyses estimating the effect of unconsidered physical contributions on the trajectories have been made, which include – among others – thermal and solar radiation force [7], atmospheric drag, Earth's gravity model (including solid and ocean tides), Earth's albedo radiation, solar wind, relativistic effects [1,9] or electro-magnetic interactions [9]. All of these could not explain the velocity increase. Furthermore, theories have been developed, proposing a hypothetical fifth force [10], new force-field models [11], or light speed anisotropy [12] to fit the anomaly; still none of these have been proven in other contexts. Other papers suggest issues in the modelling and implementation of navigation software and Orbit Determinations systems [13], based on Moyer [14] formulation.

This work revisits the Rosetta Earth flyby in 2005 and estimates the contributions of so far unmodelled physical effects on the anomaly, using the European Space Operations Centre's (ESOC's) new orbit determination system for deep space missions [15]. With this, many of the unclarified questions from literature can be tackled, with the purpose of finding a physical explanation of the anomaly in the case of Rosetta.

Table 1. Spacecraft Earth flybys with their closest approach date, measured velocity increment at perigee $\Delta V_p$ and asymptotic velocity increment $\Delta V_\infty$

| Spacecraft | Flyby date (DD/MM/YYYY) | $\Delta V_p$ / mm/s | $\Delta V_\infty$ / mm/s |
|---|---|---|---|
| Galileo I | 08/12/1990 | 2.56±0.02[3] | 3.92 |
| Galileo II | 08/12/1992 | -2.90±0.70[3] | -4.60 [a)] |
| NEAR | 23/01/1998 | 7.210±0.006[3] | 13.46 |
| Cassini | 18/08/1999 | -0.42±0.42[4] | -0.50 |
| Stardust | 15/01/2001 | ? | ? |
| Rosetta I | 04/03/2005 | 0.67±0.02[2] | 1.82 |
| MESSENGER | 02/08/2005 | 0.0071±0.004[4] | 0.02 |
| Rosetta II | 13/11/2007 | - | 0 [b)] |
| EPOXI I | 31/12/2007 | - | 0 [c)] |
| EPOXI II | 29/12/2008 | - | 0 [c)] |
| Rosetta III | 13/11/2009 | - | 0 [b)] |
| EPOXI III | 28/06/2009 | - | 0 [c)] |
| Juno | 09/10/2013 | - | ~0 [5] |
| OSIRIS-REx | 22/09/2017 | ? | ? |
| BepiColombo | 10/04/2020 | - | 0 [b,c)] |
| Solar Orbiter | 27/11/2021 | - | 0 [b)] |
| Lucy I | 16/10/2022 | ? | ? |
| JUICE I | 20/08/2024 | 0.10 [6] | 0.6 [b)] |
| Lucy II | 13/12/2024 | ? | ? |

[a)] Drag due to low altitude
[b)] Internal investigation based on original flyby analysis
[c)] Distant flybys larger than 15,000 km distance

## II. ROSETTA EARTH FLYBY

### A. *Flyby trajectory*

After its launch on 02 March 2004, Rosetta followed an interplanetary trajectory, including in total three Earth (2005, 2007, 2009) and one Mars flyby (2007) to reach the mission's target 67P/Churyumov-Gerasimenko. The first of the Earth flybys, conducted on 04 March 2005, is the main topic of this work and will be presented in more detail.

Starting from a slightly eccentric initial heliocentric orbit (aphelion: 1.09 AU, perihelion: 0.89 AU), the main purpose of the Earth flyby was to increase speed (3.797 km/s) with a slight direction change (8.106°) to target the upcoming Mars flyby. Fig. 1 shows the geocentric trajectory around perigee passage. The spacecraft approaches Earth from the nightside in a highly inclined orbit, passes by Earth at a closest approach of 1956 km above the surface on the dayside at 22:10:18 (TDB), 11 mins thereafter flying just 2.4° off the subsolar point, then departs approximately in the direction of the Moon. The osculating orbital parameters at perigee epoch in ICRF are:

- Epoch: 2005-03-04T22:10:18.3027 TDB
- Perigee distance: 8331.7981 km
- Velocity at Perigee: 10.517049 km/s
- Semi-major axis: -26704.0701 km
- Eccentricity: 1.3120048
- Inclination: 144.931532°
- R.A. of Ascending Node: 170.17250°
- Argument of Perigee: 143.09909°
- Asymptotic Velocity: 3.8634932 km/s

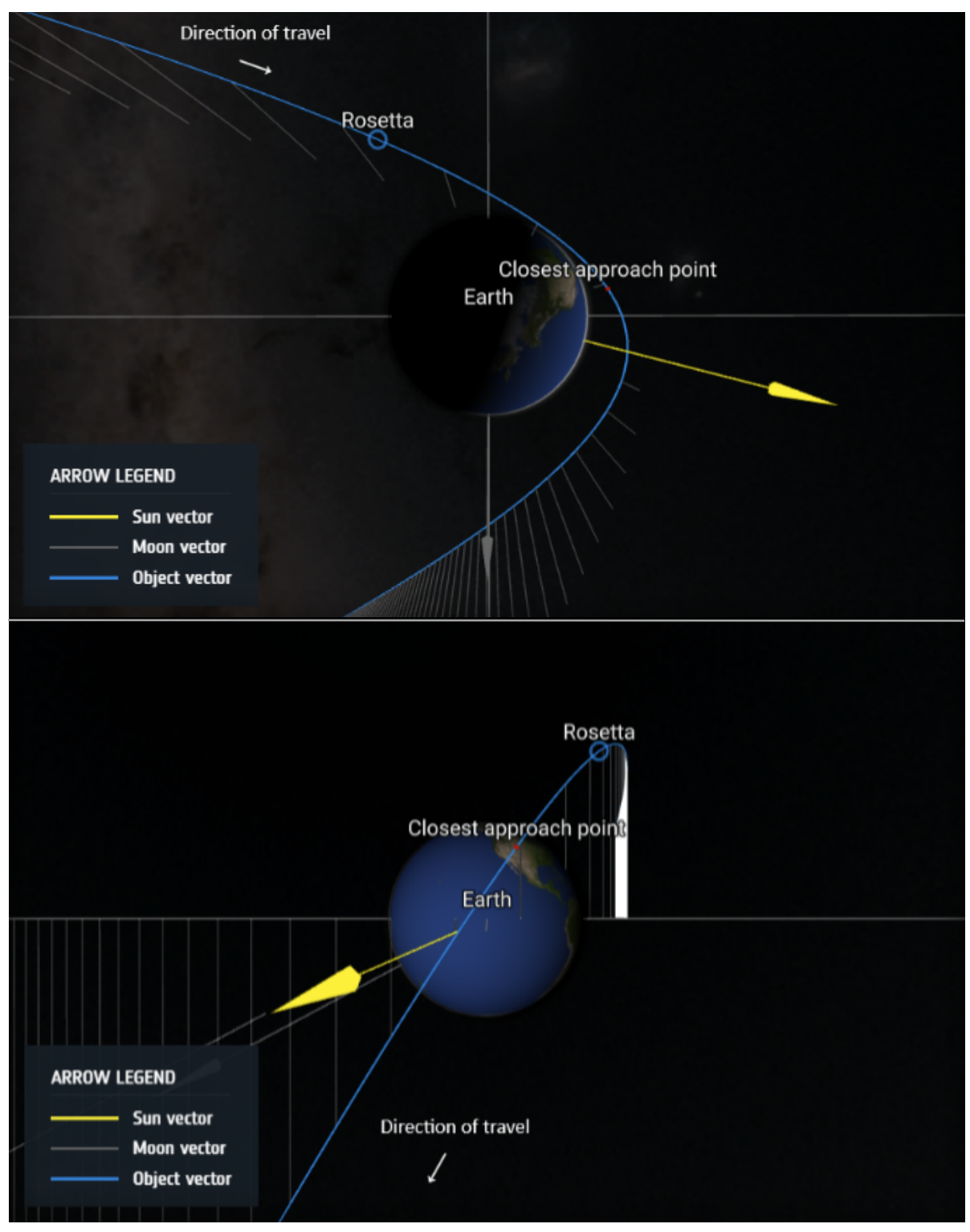

Fig. 1. Rosetta Earth flyby trajectory on 2005-03-04, Rosetta marking (blue circle) 30 min before closest approach (red), top: equatorial X/Y-plane, bottom: Y/Z-plane; visualisation from [16]

### B. *Rosetta Spacecraft Characteristics*

Rosetta is a 3-axis stabilised spacecraft with four reaction wheels that routinely need to be offloaded, on average every 5 days. These wheel offloadings (WOLs) are compensated by thrusters, thus ideally being purely torque, still marginal perturbations up to about 1 mm/s are seen in the orbit and calibrated in the orbit determination. The calculated mass at the time of the flyby is 2895.21 kg. The spacecraft consists of two solar arrays, attached along the Y-axis of the body (cf. Fig. 2), with a total area of 64.62 $m^2$. The central body has dimensions of 2.1 m x 2.0 m x 2.575 m (X/Y/Z in spacecraft frame). Majority of the spacecraft is covered with highly absorbing MLI (black Kapton). Communication and tracking via Doppler and range is performed mostly with the 2.2 m diameter High-Gain Antenna (HGA), which is 2-axis steerable and can use S- and X-band. Otherwise, two omnidirectional Low-Gain Antennas (LGAs), one fixed on the +X/+Z edge (LGA-F) and one on the -X/-Z edge (LGA-R), as seen in Fig. 2, can communicate in S-band only and are used at low Earth distances. Rosetta's scientific instruments and navigation cameras are mainly located on the +Z deck. The Philae lander is attached during cruise phase on the -X-face (Fig. 2). The launch vehicle was attached on the -Z-face, where the remaining adapter ring is usually a cold face not illuminated by the Sun.

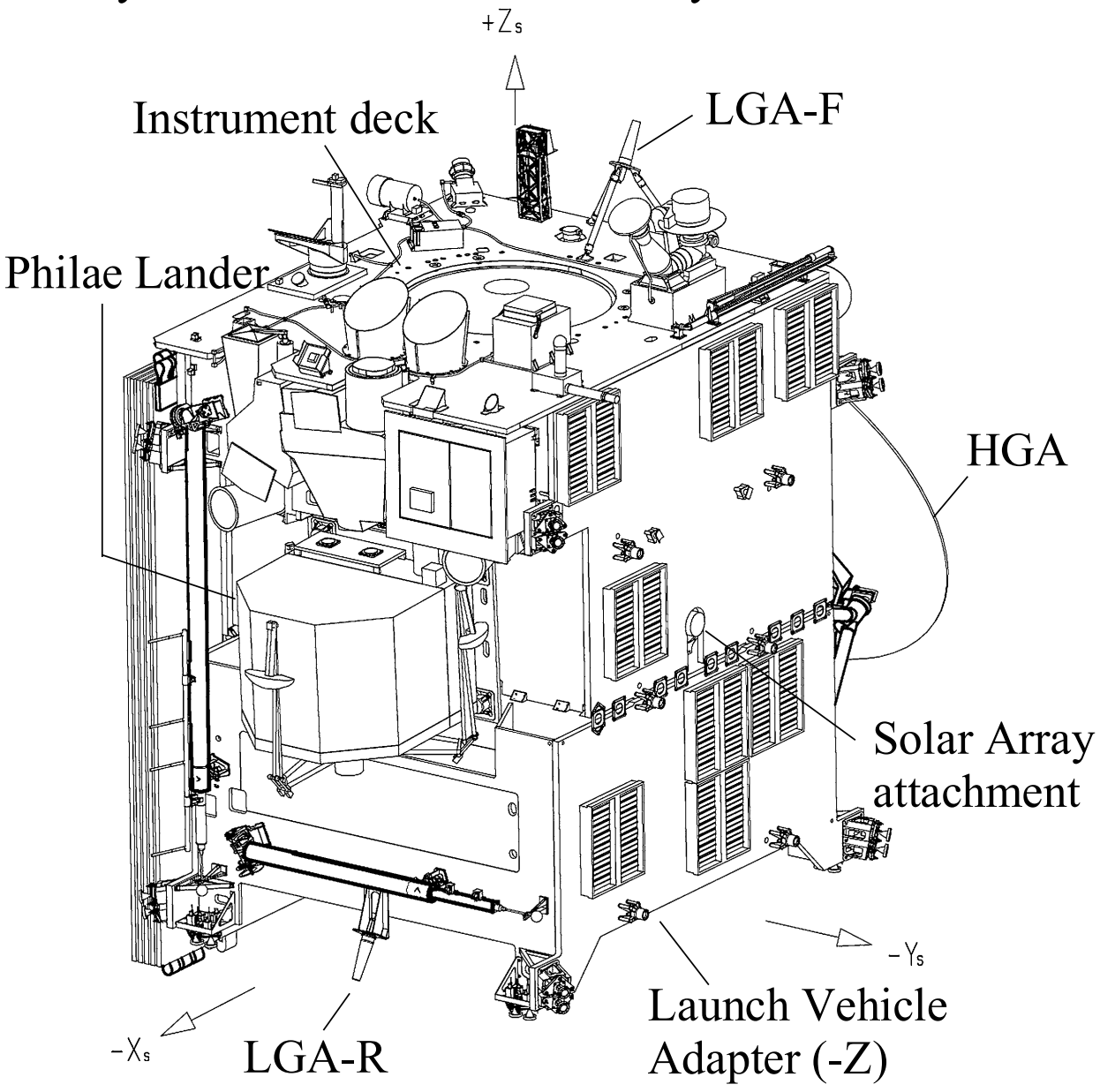

Fig. 2. Rosetta schematic from -X side in launch configuration (from [17]) with annotated features.

### C. Spacecraft constraints and guidance

Due to thermal, power, slewing limitations or instrument safety, there are several constraints on the spacecraft to consider, thus limiting the guidance around the Earth flyby. As a result, the solar array Y-axis is – per default – kept orthogonal to the Sun direction and the arrays are steered, such that their surface normal is pointing towards the Sun. Also, in the X/Z-plane there are Sun illumination constraints so that the aspect angle $\alpha$ from +X direction in this plane should stay between $-20° \leq \alpha < 25°$ at 1.0 AU. Under no circumstances Sun illuminations with $\alpha > 80°$ (on +Z face) or $\alpha < -50°$ (-Z/+X edge) are allowed to protect the lander on -X and thrusters on -Z.

During the first Earth flyby Rosetta's default guidance is used with the Sun in the spacecraft X/Z-plane. The aspect angle of the Sun from X-axis varies within 14° to 18° during ±1 h of the flyby, with an angle of 14.5° at closest approach. This means the orientation with respect to Earth varies throughout the flyby.

## III. New Orbit Determination Setup

Orbit Determination systems have improved over the years; thus, we considered it a valuable exercise to revisit the Rosetta Earth flyby anomaly from 2005 by re-running an Orbit Determination, based on the original tracking data, with ESOC's new orbit determination system for deep space missions based on GODOT [15]. This also enables, amongst other things, the incorporation of more advanced dynamics modelling, improved observation corrections, and ephemerides updates to get a better hold of the anomaly and exclude potential error sources in the original analysis. Furthermore, it is possible to further constrain the orbital velocity increment effect in magnitude, direction and time, which helps to find a physical explanation of the phenomenon.

### A. Verification of the anomaly

Replicating the setup of the original orbit determination runs, as described in [2], we find the same anomaly with the new system. As shown in Fig. 3, the post-fit Doppler residuals around perigee passage show significant signatures that are comparable to the results in [2], if using the nominal setup without an artificial $\Delta V$ at perigee.

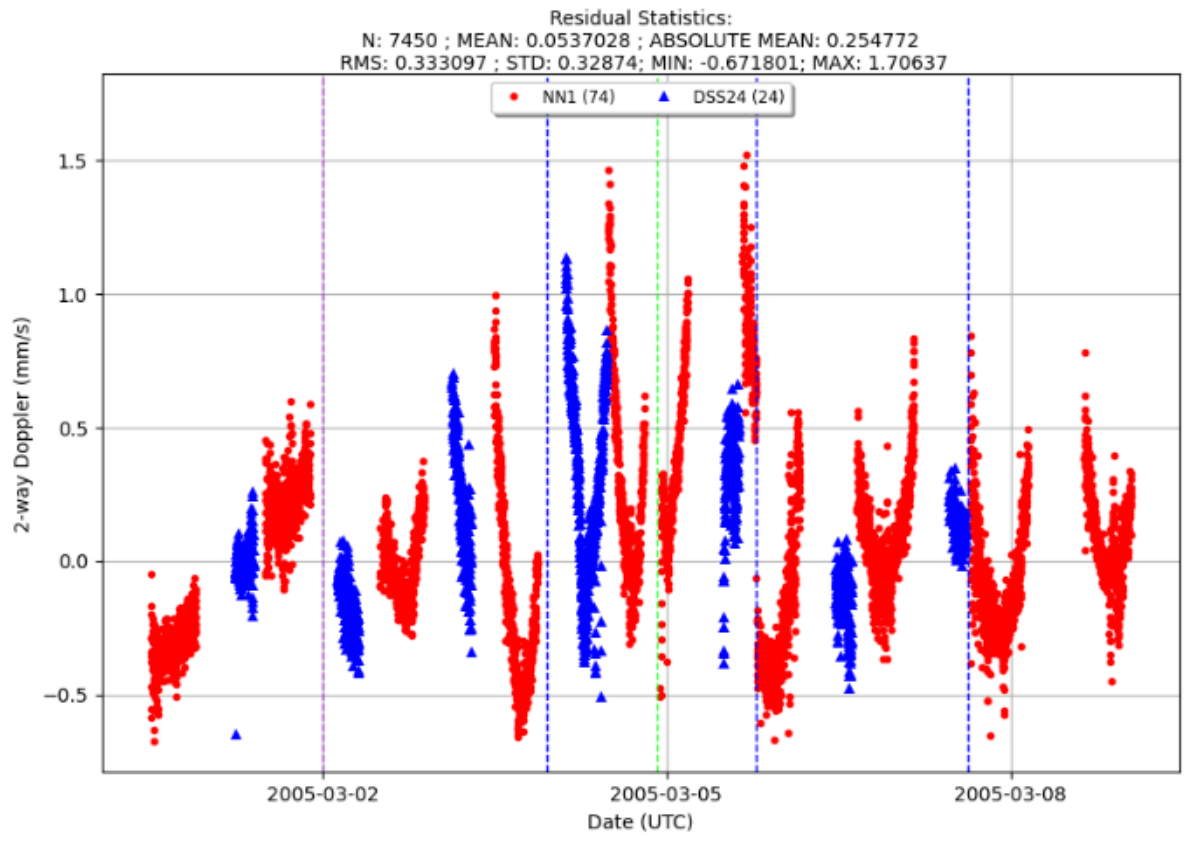

Fig. 3. Post-fit S-band Doppler residuals (2-way), modelling as in [2] with new system [15], without artificial $\Delta V$; lines: WOL (blue), outgassing (purple), perigee (green)

Assigning zero weight to the post-perigee tracking data in the orbit determination, such that it does not have any influence on the solution, shows rising offset in Doppler residuals (see Fig. 4). There is a significant jump of around 2 mm/s with the first recorded data after perigee (obtained at 22:44 UTC, 35 mins after closest approach) with a flattening over the following days, after which it reaches an offset of $(3.64 \pm 0.13)$ mm/s from pre-perigee residuals. These are comparable results to [2] showing the Rosetta orbit determination anomaly still exists when using the new system.

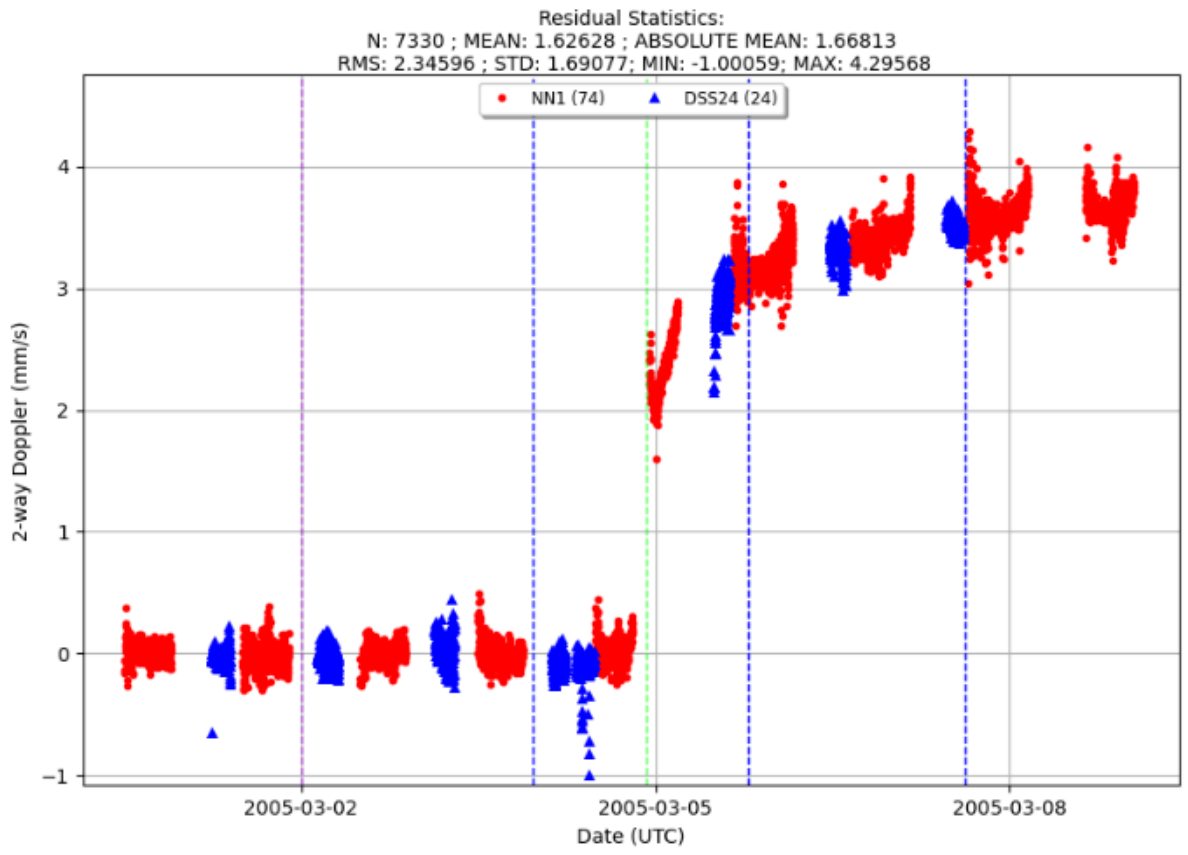

Fig. 4. Post-fit S-band Doppler residuals (2-way), with post-perigee data weighted zero; lines: WOL (blue), outgassing (purple), perigee (green)

*B. Adaptation of new observation arc and ΔV estimate*

Extending the observation arc a few more days allows for the incorporation of X-band data taken before 27 February and after 11 March. The new arc, starting on 17 February at 14:05 UTC, just after the last trajectory control manoeuvre (TCM-6), up until 12 March, includes in total 39 ground station passes. Most of the tracking data (2-way Doppler and range) were acquired from ESA ESTRACK's New Norcia 1 (NNO-1) antenna with support passes from NASA's Deep Space Network (DSN) with antennas at Goldstone (mostly DSS-24) and at Madrid complex.

Running an orbit determination with an estimated artificial prograde ΔV at perigee epoch results in a good fit with well-centred residuals and minor signatures (see Fig. 5). The estimated ΔV in this configuration is (0.655 ± 0.015) mm/s (1σ formal uncertainty), consistent within errors given in [2].

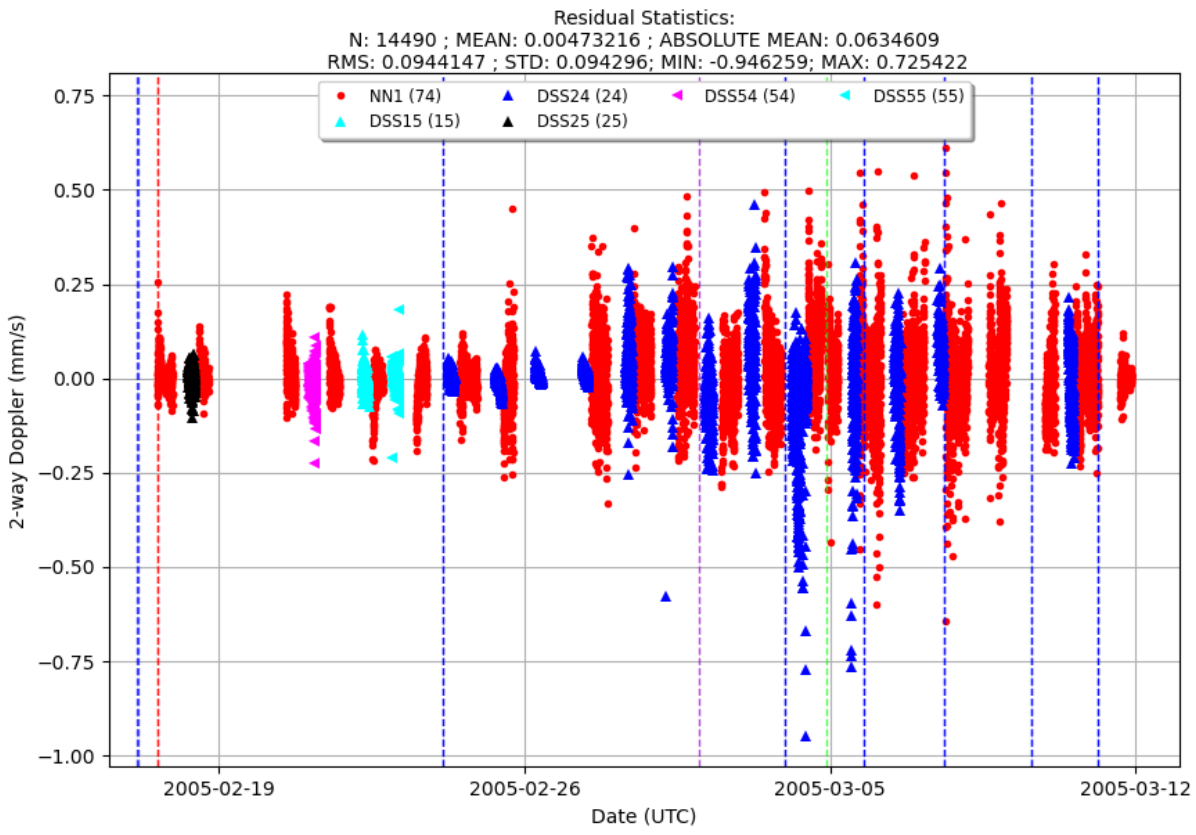

Fig. 5. Post-fit X- and S-band Doppler residuals (2-way), with artificial ΔV at perigee; lines: WOL (blue), outgassing (purple), perigee (green), TCM (red)

*C. Modification and improvement of inputs*

The Rosetta mission was flying for more than 11 years after the first Earth flyby. Since then, there have been several advancements and detailed characterisations of the spacecraft, like e.g., the LGA and HGA modelling, that have improved the orbit determination. Implementing these antenna models in the observable modelling by compensating range and range-rate by the position and movement of the antenna phase centre with respect to the spacecraft's centre of mass, helps removing several signatures seen in range and Doppler residuals, especially during slews. Some of these Doppler signatures, reaching temporarily up to 2 mm/s 2-way, were simply removed in the original analysis by deleting the data, other smaller ones (cf. Fig. 4/5) remained there. For example, the implementation of proper LGA-R causes a shift of up to 0.25 mm/s during the first pass after perigee passage, which is potentially a meaningful error source for an accurate ΔV estimation.

LGA-F was mostly used for a few days around perigee passage from 02 March at 12:05 UTC until 06 March at 16:00 UTC, only the first immediate NNO-1 pass after closest approach from 22:44 UTC until 04:25 UTC had to be taken by LGA-R for geometric reasons, otherwise the steerable HGA was used. For all further analyses, the antenna corrections were considered.

Further major improvements in orbit determination can be achieved in the dynamical modelling. The analysis in [2] used DE405 planetary ephemerides and constants, simplified spacetime and JGM3 (16x16 degrees) Earth spherical harmonics. Also, relativity effects of only the Sun were considered in the gravity model. We completely revised the dynamics setup and incorporated:

- INPOP19a planetary ephemerides, spacetime and constants [18]
- Eigen-6c 64x64 Earth spherical harmonics [19]
- IERS2010 solid (and ocean) tides model [20]
- General relativity of Sun, Earth and Moon for gravity model

Also, the orbit propagator's centre-of-integration (COI) is set to the Solar System Barycentre (SSB) outside Earth's sphere of influence (SOI), instead of previously the Sun. In Earth's SOI, propagation with respect to the Earth centre is still made. To address the theory from [13] about a formulation issue with COI switches, an independent orbit run was made with Earth as COI throughout the whole arc, but this only reproduced the same anomaly as seen before.

Including all the observable and dynamics modifications from above, results show that the orbit determination anomaly still exists in an up-to-date setup. Individual improvements change the ΔV estimates slightly (below 0.02 mm/s) each, but in both directions increasing and decreasing its size. The biggest contribution is the incorporation of general relativity of Earth in the dynamics. The overall change with all components is insignificant and amounts to about 0.01 mm/s, whilst residuals statistics slightly improve by 11% in RMS.

## IV. ANOMALOUS ΔV DETAILED ESTIMATION

With the improved observation and dynamics modelling, the anomalous ΔV around perigee can be analysed in more detail, e.g. the effects if opening the freedom in direction to allow for a not perfectly prograde velocity increment, or to allow for a time variation around perigee. For improved convergence, a short arc starting on 28 February was selected, and an initial state epoch close to the closest approach time was selected (2005-03-04T22:00 TDB) for estimation.

Estimating the ΔV as usual in simply prograde direction with the short arc gives an estimate of (0.688 ± 0.020) mm/s (1σ). An apriori sigma of 1.0 mm/s was assumed for this. Varying the apriori sigma input into the orbit determination changes the posteriori estimates of the ΔV, as seen in Table 2. For apriori sigmas greater than 0.50 mm/s, there is no significant change in the estimate and posteriori filter sigma; the estimate is stable. For small apriori sigmas (<0.25 mm/s), the estimation filter drags the solution towards slightly smaller estimates of ~0.6 mm/s, as the initial value used in all cases is 0 mm/s (assuming no anomaly as default). Anyways, the estimate is quite significantly pointing towards a positive prograde ΔV, even if not given much freedom.

Table 2. ΔV estimates and sigmas based on different apriori filter settings for prograde velocity change

| Apriori sigma (1σ) / mm/s | Posteriori estimate / mm/s | Posteriori sigma (1σ) / mm/s |
|---|---|---|
| 0.05 | 0.596 | 0.018 |
| 0.10 | 0.663 | 0.019 |
| 0.25 | 0.684 | 0.020 |
| 0.50 | 0.687 | 0.020 |
| 1.00 (nominal) | 0.688 | 0.020 |
| 2.50 | 0.688 | 0.020 |
| 5.00 | 0.688 | 0.020 |
| 10.0 | 0.688 | 0.020 |

The next step is to estimate the ΔV in three dimensions. As a default, a tangential axis in the direction of the instantaneous velocity vector is selected to calibrate the increment in prograde direction (as done before); for the other directions the orthonormal base is selected in a way that the second component is in the direction of the orbital momentum vector (describing a cross-track direction) and the third one completes the system right-handed system (describing a quasi-radial velocity direction). We call this local velocity-based base vector set the TCN frame. In this frame, the parameters can be described in a spherical representation (magnitude, angles θ and φ). Constraining apriori the ΔV magnitude again to 1.0 mm/s with direction sigmas of 90° (1σ) each, gives an estimate of ΔV = (0.695±0.195) mm/s with θ = (-0.3±89.9)° and φ = (-9.3±65.3)°. This shows that the elevation above the orbital plane θ (or cross-track direction) is basically unresolvable, as the apriori sigma matches the posteriori one, but the angle in the orbit plane φ (or radial direction) is slightly more observable with a tendency towards a slight outwards movement as seen from Earth. Nevertheless, the posteriori sigma is significantly larger than the estimated value, thus not being representative.

Another way to quantify the three-dimensional ΔV is in spacecraft frame. Projecting the TCN result into spacecraft frame based on the attitude quaternions, gives ΔV (x,y,z) = (0.436, -0.025, 0.540) mm/s, so a major contribution in the spacecraft +Z direction with a slightly smaller one in +X direction. Estimating the ΔV direction in spacecraft frame directly and varying the apriori sigma spherically (same value in each direction), leads to the results in Table 3. They show that significant ΔVs can be estimated and resolved in X- and Z-axes, with slightly better resolvability in Z-direction. The estimates in Y-direction barely reduce the posteriori sigma independent of the apriori sigma, thus not being resolvable at all. For low apriori sigmas (below 1.0 mm/s) the dynamic freedom is used towards an +Z major contribution and a slightly above half the size contribution in +X. Both components are resolvable. At a spherical apriori of 1.0 mm/s both contributions are the same and for even larger apriori setups the freedom is rather used in +X direction. However, posteriori sigmas are larger than the actual estimates for both, thus being not fully resolvable.

Table 3. ΔV estimates and sigmas based on different apriori filter settings for 3-dimensional velocity change

| Apriori sigma (spherical, 1σ) / mm/s | Posteriori estimate ± 1σ uncertainty / mm/s | | |
|---|---|---|---|
| | S/C X-direction | S/C Y-direction | S/C Z-direction |
| 0.05 | 0.296±0.044 | -0.036±0.050 | 0.516±0.030 |
| 0.10 | 0.331±0.087 | -0.040±0.099 | 0.572±0.053 |
| 0.25 | 0.356±0.213 | -0.040±0.249 | 0.583±0.124 |
| 0.50 (nominal) | 0.402±0.403 | -0.035±0.498 | 0.561±0.233 |
| 1.00 | 0.514±0.680 | -0.026±0.994 | 0.498±0.397 |
| 2.50 | 0.697±0.986 | -0.049±2.460 | 0.392±0.615 |
| 5.00 | 0.773±1.122 | -0.209±4.813 | 0.337±0.819 |
| 10.0 | 0.840±1.308 | -0.765±8.918 | 0.260±1.216 |

An additional check was made by comparing two independent orbit solutions, one based on pre-perigee tracking data and the other with post-perigee data alone.

The orbits' difference can be seen in Fig. 6. They differ mostly by a radial component velocity, as seen from Earth, of around 2 mm/s in the outgoing asymptote, which directly translates into a range-rate jump. Vice versa, the incoming asymptote miss-matches with respect to a post-perigee solution by around 2 mm/s, as well. There is an additional linear drift visible pre-perigee that is due to solar radiation pressure estimated slightly differently before the flyby. There are additional jumps visible in Fig. 6 that are due to estimated WOLs. Interesting to note is a near constant -0.7 to -0.9 mm/s along-track difference of both solutions within a day before and after flyby, where there is no other dynamic event.

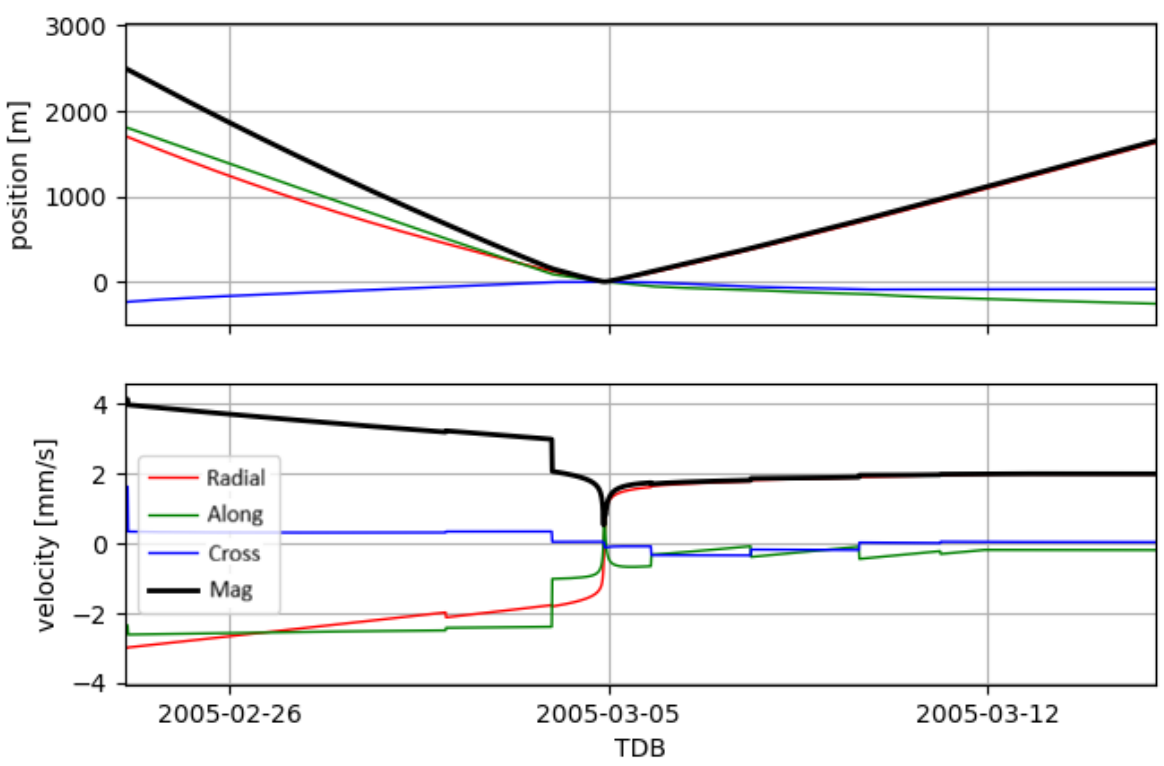

Fig. 6. Orbit position and velocity difference between independent pre- and post-perigee solutions

Attempts have been made to estimate the time of the ΔV around perigee within the observation gap from 19:20 – 22:44 UTC, but no significant results that improve the orbit determination could be obtained. Estimating a time offset with a 3-dimensional ΔV does not lead to resolvable results. Only in the prograde case, when leaving the freedom fully open (true anomaly of 90°, 1σ), the estimation filter moves the velocity increment slightly to 3 mins before closest approach, or (-12.9 ± 68.0)° in true anomaly, which is still barely observable. The ΔV estimate rises then to (0.692 ± 0.059) mm/s.

## V. ADDITIONAL PHYSICAL EFFECTS

During the observation gap of the two ground passes around perigee passage, there could be other, yet unconsidered, physical effects that might influence the trajectory of Rosetta at its Earth flyby. Several remote sensing on-board instruments participated in the flyby operations, whose data are publicly available by now in the ESA Planetary Science Archive (PSA). The idea is to link a dynamical effect to scientific results or other housekeeping data (cf. VI.) seen.

A first candidate for additional dynamics is the Earth's radiation and electromagnetic environment. On the latter, Rosetta's fluxgate magnetometer (RPC-MAG) [21] provided interesting results during the flyby in March 2005. As the spacecraft came out of deep space from almost aligned to the nightside magnetotail, then entering magnetopause early on 04 March and passing the dayside plasmasphere of Earth at perigee passage later that day, and last exiting the magnetosphere on 05 March at 02:02 UTC and crossing the outbound bow shock later that day, it gave some opportunity to investigate the Earth's magnetic environment. Interestingly, [21] reports a large unexplained measured magnetic field difference with respect to the reference model at around 22:04 UTC – shortly before the flyby – lasting for a few minutes. As described by [22], Rosetta travels through parts of the radiation belts with an expected peak flux for protons within 30 mins around closest approach to Earth. The predicted electron flux is distributed around 2 hours around the flyby. The radiation pressure (<$10^6$ protons / (cm$^2$·s) for 30 MeV, <$10^7$ electrons / (cm$^2$·s) for 5 MeV) should not cause accelerations larger than $10^{-11}$ m/s$^2$ on the spacecraft (using parameters in II. B.), which is insignificant for the flyby anomaly.

Other effects in Earth's environment are the albedo and infrared radiation. Rosetta is moving on the dayside of Earth around closest approach time. Due to the low altitude (~0.3 $R_E$), Earth should cover a significant proportion of the field of view as seen from the spacecraft. Assuming a 340 W/m$^2$ irradiation from Earth (about a fourth of the incoming solar radiation), this would cause an acceleration of $2.5·10^{-8}$ m/s$^2$ on Rosetta's solar arrays straight-on if fully absorbed. This is one order of magnitude larger than estimated in [9] because of the high area-to-mass ratio, but this is still too small to explain the anomaly as seen by [1] in the order of magnitude of ~$10^{-4}$ m/s$^2$. For Rosetta anyhow, an acceleration of $10^{-6}$ m/s$^2$ over the course of 10 mins would be sufficient to explain the anomaly.

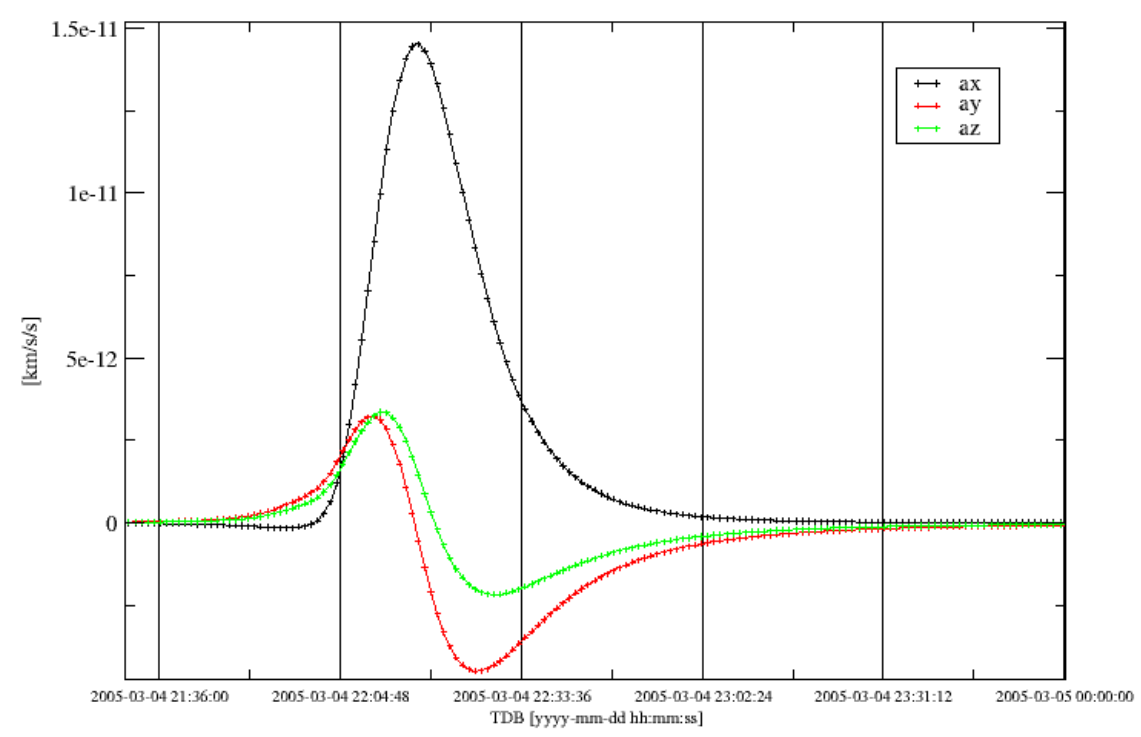

Fig. 7. Earth albedo acceleration computed for Rosetta accurate model in ICRF around flyby

Incorporating an Earth radiation model as described in [7] on the described properties of Rosetta with reconstructed orbit and attitude information, HGA and solar array articulations, leads to the computed accelerations as seen in Fig. 7. Its magnitude peak reaches $1.5 \cdot 10^{-8}$ m/s$^2$ (as solar arrays are not perfectly orthogonal to Earth), mostly in sunward direction. Running an orbit determination with this additional acceleration reduces the artificial prograde ΔV estimate slightly to (0.650 ± 0.015) mm/s, but not removing it, as expected.

## VI. Outgassing Hypothesis

Further analyses of the albedo radiation force show that the effective magnitude needs to be ~80 times larger to fully explain the flyby anomaly. This raises the question whether Earth's illumination could have triggered an outgassing on cold spacecraft faces with potential sublimating ice.

Rosetta is by its design thermally sensitive to illumination on the -Z and -X faces. The launch vehicle adapter (-Z) is not thermally protecting the NTO (oxidizer) tank; also, there are several thrusters located on the -Z face. On the -X face, the Philae lander, several instrument radiators and two star-trackers make it a thermally sensible face and need strictly to be kept cold (cf. II. C.). Both conditions were originally not checked with respect to Earth's albedo radiation. Outgassings on these faces could provide a velocity increment matching in direction with the flyby anomaly estimates (cf. IV).

### A. *History of outgassings*

In the mission's lifetime there have been multiple confirmed and additionally suspected outgassings, especially during the early orbit phase and the first years in-flight. In an internal archival investigation, we were able to link all recorded instances of outgassing up until the Earth flyby to illuminations from the Sun on -Z face during slews. Measured ΔVs vary with illumination angle and duration but reach magnitudes from 0.05 to 1.9 mm/s projected in the line-of-sight as seen from Earth. Vice versa, all illuminations of -Z during this period, some of which reaching an aspect angle $\alpha = -50°$ (cf. II. C.) for a few hours during slews, caused a reported outgassing. No occurrences of -X illuminations were possible during this time. Only much later, there was a reported outgassing during a short -X illumination in February 2010 at 1.58 AU Sun distance with +0.05 mm/s in Earth's line of sight.

Only shortly before the Earth flyby, there was a suspected instance of an outgassing during a 180° attitude flip on 02 March 2005 with a ΔV of around 0.1 mm/s. The spacecraft had a similar condition of a -Z illumination at -50° aspect angle on 23 May 2005, the first one after the flyby, but this time no ΔV could be measured.

### B. *Albedo illumination conditions*

As the spacecraft guidance is nearly fully constrained with respect to the Sun (II. C.) and the spacecraft is moving in an eccentric trajectory around Earth (II. A.), the viewing conditions of Earth from the spacecraft are constantly changing during the flyby. Fig. 7 shows the angles between the spacecraft frame axes and Earth direction in the hours around closest approach. Rosetta's Y-axis stays nearly orthogonal to Earth direction (within 8°), whilst Earth is shining on different faces in the X/Z-plane. The Earth's terminator crossing happens at 21:58 TDB, 12 mins before closest approach (cf. Fig. 1). The -X face of the spacecraft is visible from Earth from 21:52 TDB up until far after the flyby at 00:20 TDB but being nearly fully parallel (within 4°) to Earth at 22:17 TDB, shortly after closest approach. The -Z face starts being illuminated at this exact time and lasts for many hours thereafter, but the illumination is then limited by the geometric distance and phase angle of Earth.

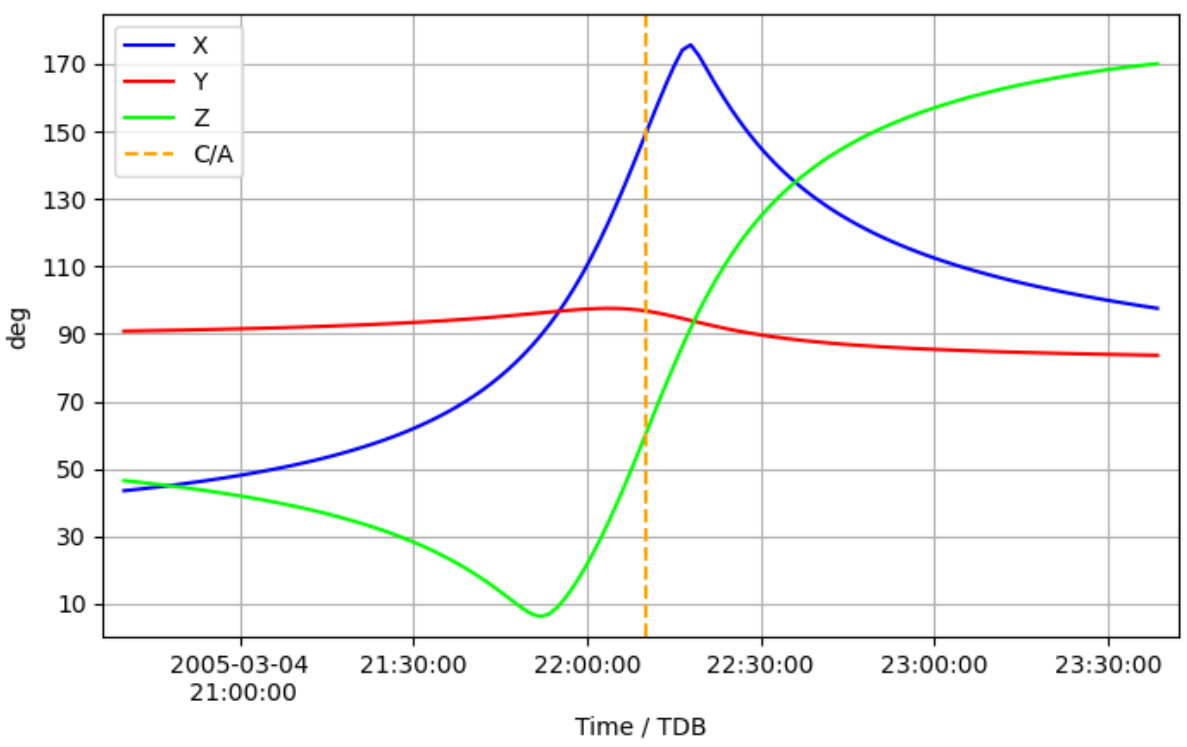

Fig. 7. Angles of spacecraft face normals towards Earth around flyby closest approach (C/A).

To quantify this effect over time, the albedo radiation pressure was computed in spacecraft frame, as shown in Fig. 8. It is equivalent to the energy received per area and time divided by the speed of light. The -X face albedo illumination has a major contribution and peaks around the closest approach; the -Z contribution is smaller and peaks around 22:28 TDB after the flyby. The absolute values of the irradiance pressure are about one magnitude smaller than the corresponding solar radiation pressure at 1.0 AU (~$4.5 \cdot 10^{-6}$ N/m$^2$).

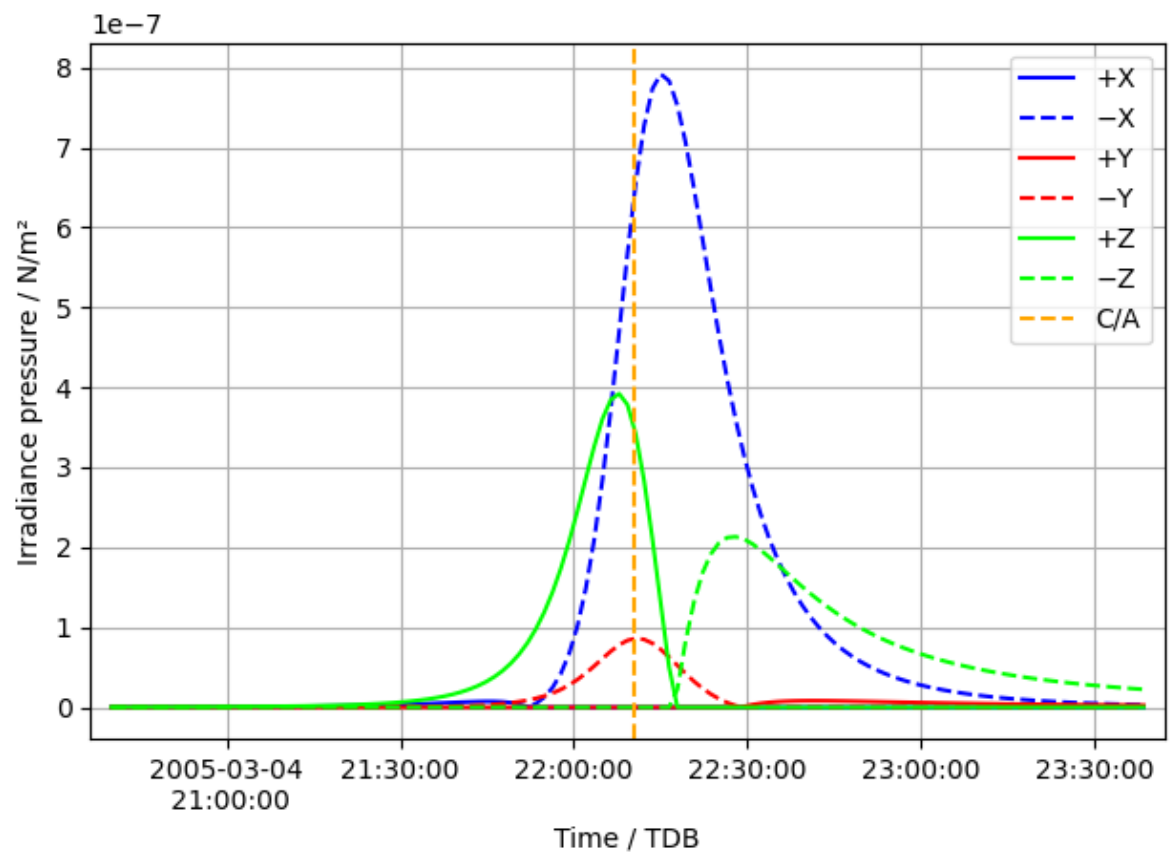

Fig. 8. Earth albedo radiation pressure on spacecraft faces around flyby closest approach (C/A).

### *C. Additional spacecraft torques*

To have a more detailed look into this theory, the reaction wheel levels of the spacecraft can be analysed. For an outgassing at one of the spacecraft faces, there would be an additional torque applied on the body if not aligned perfectly in the centre-of-mass axis. Therefore, the original telemetry data of Rosetta's first Earth flyby were reprocessed and the onboard measured wheel levels were obtained. On the other hand, the predicted reaction wheel levels based on the precise spacecraft model and commanded attitude, including slews, WOLs, solar radiation pressure and a gravity gradient model were extracted from the original flight dynamics commanding system computations.

The predicted and measured angular momentum on the four reaction wheels were then projected into the spacecraft frame and differenced. The resulting excess angular momentum compared to the model around the closest approach to Earth is shown in Fig. 9. The values begin with an offset. This is due to accumulated unmodelled momentum since the last WOL. Apart from that, a significant signature can be seen within 30 mins around closest approach, especially around Y-axis. There are two major bumps: one starting 10 mins before closest approach and continuing up until shortly after, the other starting 15 mins afterwards. The total angular momentum increase is around 0.4 Nms in that axis. Around Z-axis there is only a small change shortly before perigee, around X-axis there is temporary build-up of angular momentum, which is fully compensated after the flyby. This behaviour could be explained by a small mismodelling of the gravity gradient torque.

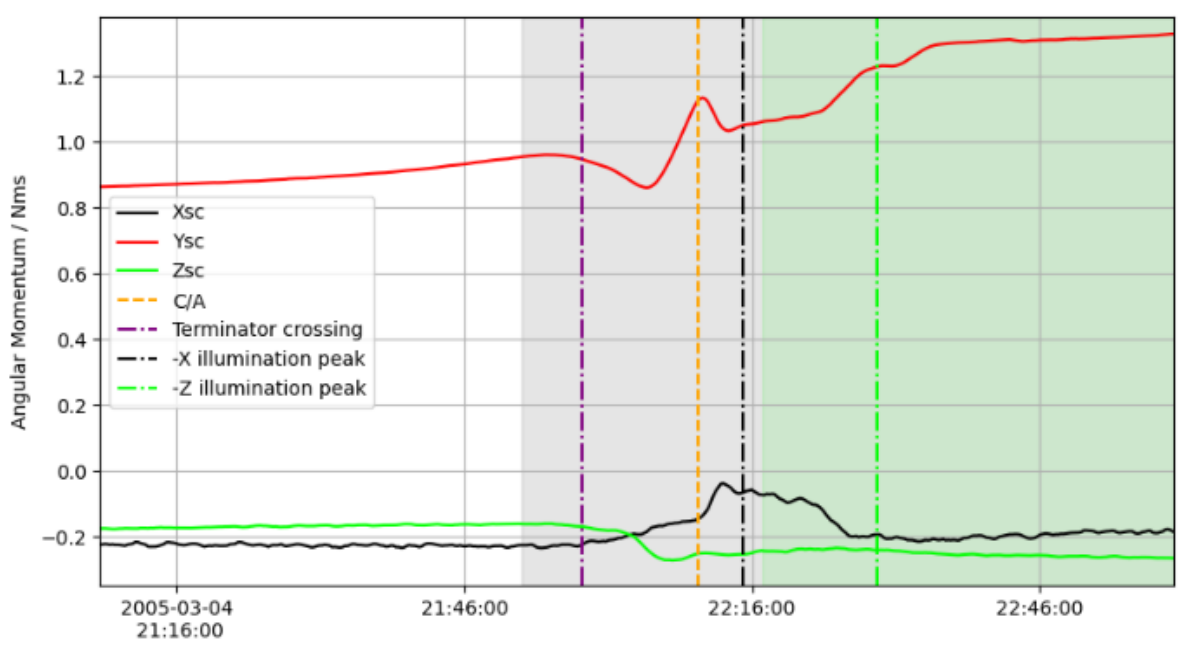

Fig. 9. Reaction wheel additional angular momentum around Earth flyby projected in spacecraft frame; illuminations shaded (-X: black, -Z: green)

Differentiating the angular momentum with respect to time leads to the torque applied to the spacecraft. Any additional torque compared to modelled torque acts as excess external torque. Fig. 10 shows both the modelled and measured torque (above) and their difference (below) in the spacecraft frame. The two signatures of additional torque around Y-axis, become even more prominent than before. For the additional torque plot, the measurements have been slightly smoothed with an average filter over five data points.

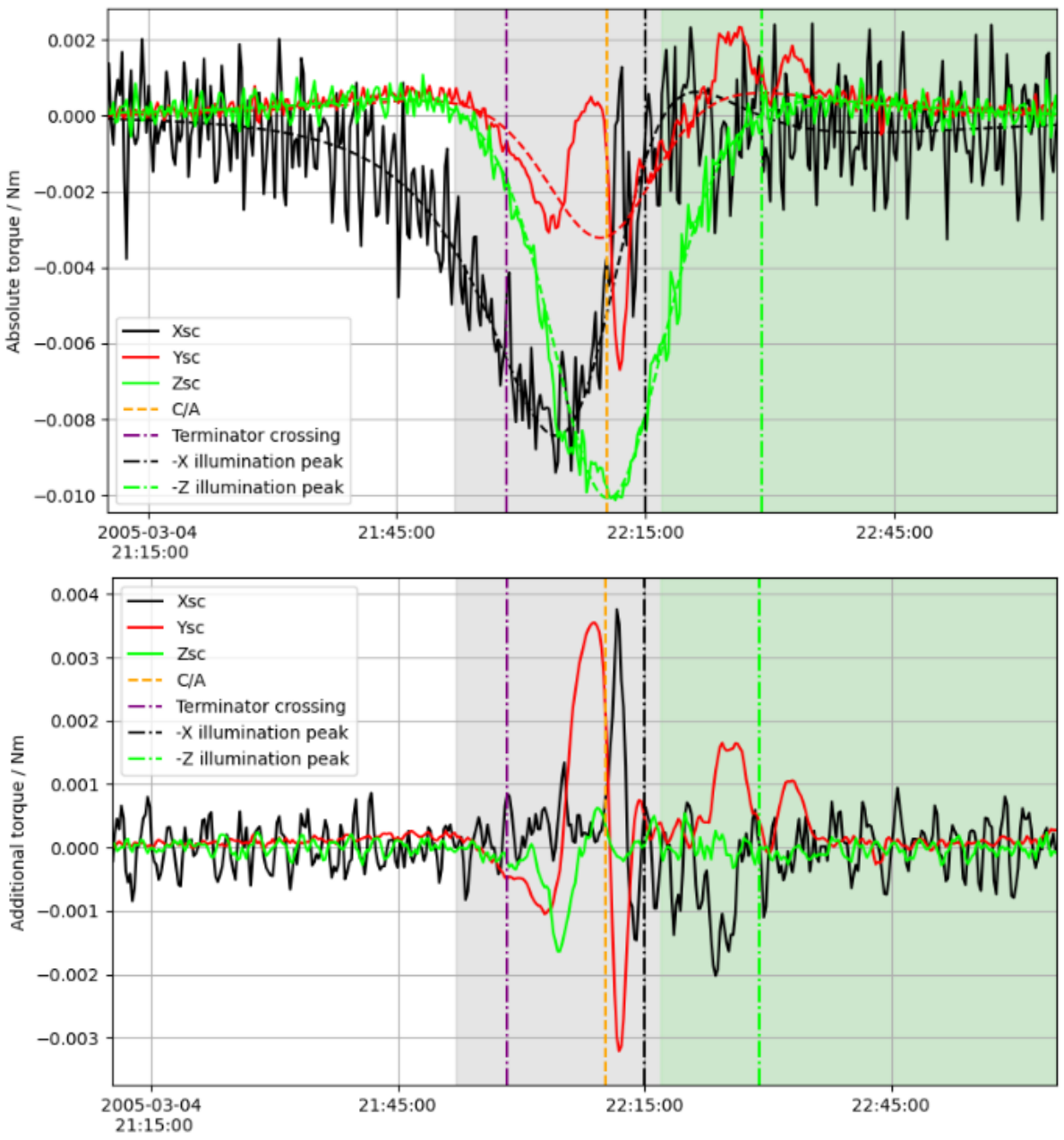

Fig. 10. Torque derived from angular momentum projected in spacecraft frame; above: data measured onboard (solid) vs. model (dashed), below: torque difference to model

The signatures start just after the terminator crossing, 12 mins before closest approach, with a small additional (negative) torque around Y-axis and then Z-axis. The torque around Y-axis changes sign 5 mins before closest approach, when Earth albedo illumination on -X face becomes predominant. The major positive peak in torque of up to 3.5 mNm lasts up until closest approach, flips sign again, but for a more limited period than the upward peak. During this time, a short peak around X-axis can also be seen, but this could be due to noise in the telemetry data. Two additional smaller peaks in torque at Y-axis can again be seen at 15 mins and 23 mins, reaching 1.6 and 1.1 mNm, respectively.

These signatures occur in time just after Earth albedo illumination on -Z face starts and peaks in intensity (cf. Fig. 8). For the main torque signature shortly before closest approach, the corresponding illumination at -X face is even more than double as high. The Earth radiation pressure alone would be too small to explain these torques. However, they match in direction: a force applied on -X face, roughly orthogonal to Y-axis and towards the -X/+Z edge would cause a positive torque around Y-axis. The same is true for a force applied on -Z face towards the -X/-Z edge, which are the viewing angles of Earth during the peak torque times.

Based on an estimated $\Delta V$ of ~0.6 mm/s, this would require an impulse of 1.74 Ns for the mass of Rosetta. With an effective lever arm of 0.5 m (half the size of the length from center-of-mass), this would result in an estimated excess angular momentum from an outgassing of ~0.87 Nms, which is double as high as it was measured onboard. With an effective lever arm around half the size, this would match the observations.

*D. Comparison to other Rosetta Earth flybys*

There have been two further Earth flybys of Rosetta. The second flyby in Nov 2007 was at a distant altitude above the surface of 5295 km; the third one in Nov 2009 was much more comparable to the first with an altitude of 2480 km. Nonetheless, no anomalous $\Delta V$ could be detected during both flybys, whereas [3] predicted a $V_\infty$ change of about 1 mm/s.

Reconstructing the orbit and attitude information of these flybys, allows for an analysis of the orientation of the Rosetta with respect to Earth, as it has been done for the first Earth flyby. Based on the angles of the spacecraft faces towards Earth and the flyby geometry, the Earth albedo radiation pressure has been computed for the second and third Earth flyby of Rosetta (cf. Fig. 11). Noticeable is the lack of significant -X and -Z face illuminations, compared to the first flyby (Fig. 8). In both flybys the illuminations of each face are more than 4 times smaller than those observed before, which makes it more unlikely to cause outgassings in these instances.

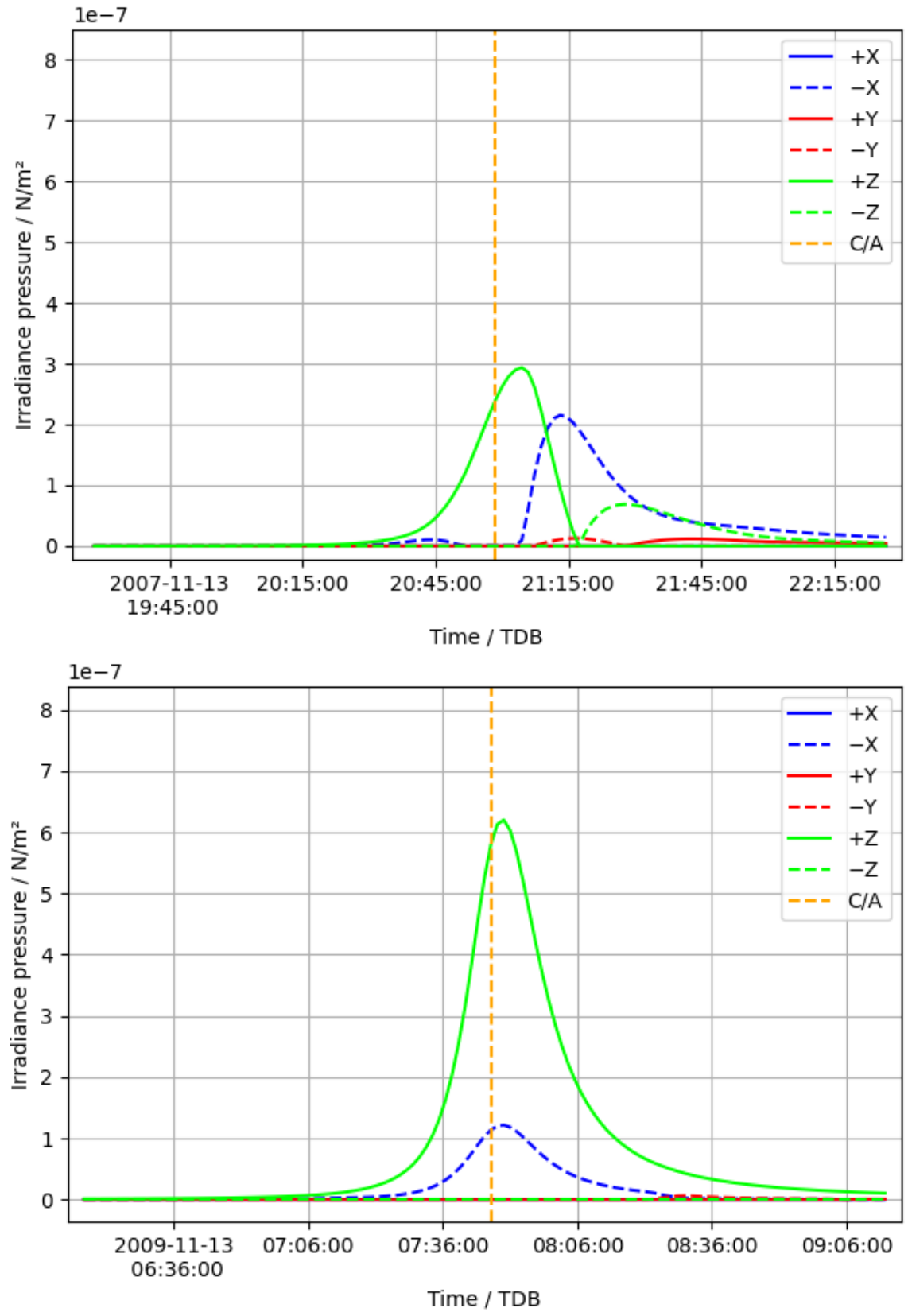

Fig. 11. Earth albedo radiation pressure on spacecraft faces at Rosetta Earth flybys 2 (13 Nov 2007) and 3 (13 Nov 2009)

## VII. DISCUSSION

During the revisit of Rosetta's first Earth flyby, several explanations for the observed flyby anomaly have been investigated. Different physical effects have been analysed in more detail, among which the Earth ephemeris errors, improved gravity spherical harmonics of Earth and relativity corrections could not be identified to be the root-cause of the anomaly. Effects of Earth solid and ocean tides have been also analysed. The maximum additional acceleration turned out to be around $10^{-7}$ m/s$^2$ for the solid and $3\cdot10^{-8}$ m/s$^2$ for the ocean tides, but both having only a minor effect (in both direction and magnitude) on the asymptotic velocity change.

Several other attempts have been made additionally. For example, estimating the Earth's gravitational parameter or its ephemeris could not solve the anomaly. The latter has been proposed initially by [2], but the old orbit determination software at ESOC did not allow for an ephemeris variation of Earth for the observation modelling, only for dynamics. With the new software's description, this is possible. However, no reasonable Earth ephemeris variation allows for the anomaly to vanish. The results described in [2], which fixes the anomaly by moving Earth in -1.4 mm/s, cannot be confirmed with simultaneous application of Earth's ephemeris shift in dynamics and observation modelling. Another idea proposed in [2] was to use a relativistic formulation in the geocentric relativistic frame of reference (GCRS). We have used an improved, but still low fidelity model in GCRS frame, but it did not fix the flyby anomaly, as well.

A three-dimensional $\Delta V$ analysis shows a consistent velocity increment in the spacecraft +X/+Z quadrant, with the orbit determination preferring slightly more the +Z direction due to better observability. On the other hand, a time estimation of the $\Delta V$ slightly prefers a time shortly before closest approach, but within large error margins. In direction and time this matches with the observations on additional torques measured onboard on the reaction wheels a few minutes before closest approach.

These effects are compatible with an Earth albedo induced outgassing on Rosetta's cold faces -X/-Z causing a push in the opposite direction. The fact that every -Z illumination caused an outgassing with $\Delta V$ measured in the entire mission life before the first Earth flyby, together with the observation that no outgassing was seen in the instances of similar illuminations for multiple years after the flyby, including the time of the following flybys, points towards a potential removal of (most of) ices on these faces during the first Earth flyby. As illumination conditions were also less favourable for outgassings on the following Rosetta Earth flybys, these both findings provide an explanation that no anomalous velocity increase could be seen for the two latter flybys.

From the analysis it cannot be fully distinguished whether a -X or -Z face albedo exposure and outgassing is more likely to cause the anomaly. The -Z option is the better observable one. However, if giving the freedom to the estimation filter, an ΔV in similar magnitude caused from -X outgassing can explain the anomaly, as well. Possible is an overlap of both contributions. Following the estimation from [24], the necessary minimum mass of sublimating water ice would be 1.73 to 1.66 g, considering a surface temperature at -X or -Z face of 0 to 20 °C on radiators (internal report on thermal radiation) respectively, for a maximum outgassing velocity of 1003 to 1049 m/s. Earlier studies estimated -170 °C on the cold spacecraft faces without albedo radiation [23], which corresponds to 2.80 g minimum sublimating water at a maximum 619 m/s. However, the mass could be considerably larger if the outgassing direction is not aligned with the surface normal.

There have been similar instances of albedo induced outgassings on spacecraft during planetary flybys. A comprehensive study has been made on the instance of BepiColombo on its second Venus flyby by [24,25], where there was a similarly sized flyby anomaly (~0.5 mm/s) around closest approach, which could be traced back to an outgassing on one of the spacecraft's main radiators due to Venus' albedo. Based on on-board torque and accurate accelerometer data of the ISA payload, they could estimate the direction of the force acting on BepiColombo. Despite a lack of appropriate accelerometer data for Rosetta, a connection of both instances can be still made through the particular flyby geometries with a day-side closest approach and the cold spacecraft faces aligned towards the planet. Also, the required sublimating ice mass is fairly similar (~2 g for BepiColombo).

Also, in the case of the JUICE Lunar-Earth Gravity-Assist [6], the observed ΔV of ~0.5 mm/s at closest approach to the Moon could be linked to an outgassing based on onboard instrument data. Additional torques of around 6 mNm, similar in magnitude to the Rosetta instance, have been observed at closest approach.

To facilitate independent analysis, we plan to provide a standalone data package of the Rosetta Flyby to the interested community.

## VIII. Conclusions

There have been several attempts in literature to explain the Earth flyby anomaly with unconsidered physical effects to consider for the spacecraft around closest approach. In this study, an improved dynamics and estimation setup could prove that various proposed ideas from literature do not explain the Earth flyby anomaly in the case of Rosetta's first Earth flyby in 2005 in practice. On top of that, observable miss-modelling and other orbit determination software implementation issues could not be identified as the reason for the anomaly.

A detailed orbital analysis was made to constrain the unexplained velocity increment in direction and time with improved dynamics. Overall, there is a strong correlation between Earth's albedo illumination and the external torques observed onboard Rosetta. While radiation pressure alone cannot explain the magnitude of the observed torques and anomalous spacecraft acceleration, outgassing triggered by this illumination is the most evident reason based on Rosetta's outgassing history to amplify the effect and explain the observed flyby anomaly.

## IX. Acknowledgements

The authors would like to thank flight dynamics colleagues (S. Kielbassa, M. Müller, V. Companys), especially the ORB team (B. Godard, F. Castellini, G. Bellei, T. Araújo), former members (T. Morley) and staff from other sections (S. Lodiot, M. Lanucara) for the archival work, technical support, documentation provision and mentorship. This work was made possible by ESA's Graduate Traineeship programme.

## X. References

[1] P.G. Antreasian and J.R. Guinn, "Investigation into the unexpected Delta-V increases during the Earth Gravity Assist of Galileo and NEAR", *AIAA 98-4287*, 1998.

[2] T. Morley and F. Budnik, "Rosetta Navigation at its first Earth Swing-by", *ISTS 2006-d-52*, pp. 593–598, 2006.

[3] J.D. Anderson, J.K. Campbell, J. Ellis and J.F. Jordan, "Anomalous Orbital-Energy Changes Observed during Spacecraft Flybys of Earth", *Phys. Rev. Lett.*, Vol. 100, 091102, 2008.

[4] B. Jouannic, R. Noomen, and J.A.A. van den Ijssel, "The Flyby Anomaly: An Investigation into Potential Causes", *ISSFD 2015*, 2015.

[5] P. Thompson, M. J. Abrahamson, S. Ardalan and J. Bordi, "Reconstruction of Earth flyby by the Juno spacecraft", *AAS 14-435*, 2014.

[6] T. Syndercombe, F. Castellini, F. Budnik and R. Meles, "The JUICE Navigation Campaign for the first combined Lunar-Earth Gravity Assist", *SpaceOps-2025*, ID 286, 2025.

[7] M. Terauchi, I. Kim, T. Hanada and J. C. van der Ha, „Effect of thermal radiation force for trajectory during swing-by", *ISST 2008-d-60*¸2008.

[8] T. Kato and J. C. van der Ha, "Precise modelling of solar and thermal accelerations on Rosetta", *Acta Astronautica*, Vol. 72, pp. 165–177, 2012.

[9] C. Lämmerzahl, O. Preuss and H. Dittus, „Is the Physics Within the Solar System Really Understood?", In: *Lasers, Clocks and Drag-Free Control; Astrophysics and Space Science Library*¸Vol. 349, 2008.
[10] L. Acedo, "The flyby anomaly: a multivariate analysis approach", *Astrophysics and Space Science*, Vol. 362-42, 2017.
[11] W. Hasse, E. Birsin and P. Hähnel, "On force-field models of the spacecraft flyby anomaly", arXiv:0903.0109, 2009.
[12] R. T. Cahill, "Resolving spacecraft Earth-flyby anomalies with measured light speed anisotropy", *Progress in Physics*, Vol. 3, pp. 9–15, 2008.
[13] S. G. Turyshev and V. T. Toth, "The Puzzle of the Flyby Anomaly", *Space Sci Rev*, Vol. 148, pp. 169–174, 2009.
[14] T. D. Moyer, "Formulation for Observed and Computed Values of Deep Space Network Data Types for Navigation", *Deep Space Communications and Navigation Series*, Monograph 2, 2000.
[15] F. Castellini, B. Godard, D.A. Dei Tos, R. Mackenzie and F. Budnik, "ESOC's new orbit determination system for deep space mission operations", *ISSFD 2022*, 2022.
[16] ESA NEO Coordination Centre, Flyby Visualisation Tool, 2026.
[17] D. Koschny, V. Dhiri, K. Wirth, J. Zender, R. Solaz, R. Hoofs, R. Laureus, T.-H. Ho, B. Davidson and G. Schwehm, „Scientific Planning and Commanding of the Rosetta Payload", *Space Science Reviews*, Vol. 128, pp. 167–188, 2007.
[18] A. Fienga, P. Deram, V. Viswanathan, A. Di Ruscio, L. Bernus, D. Durante, M. Gastineau and J. Laskar, „INPOP19a planetary ephemerides", *Notes Scientifiques et Techniques de l'Institut de mécanique céleste*, Nr. 109, 2019.
[19] R. Shako, C. Förste, O. Abrikosov, S. Bruinsma, J.-C. Marty, J.-M. Lemoine, F. Flechtner, H. Neumayer and C. Dahle, "EIGEN-6C: A High-Resolution Global Gravity Combination Model Including GOCE Data", In : *Observations of the System Earth from Space – CHAMP, GRACE, GOCE and future missions*, pp. 155-161, 2013.
[20] G. Petit and B. Luzum, "IERS conventions (2010)", *IERS Conventions Centre,* IERS Technical Note No. 36, 2010.
[21] K.-H. Glassmeier, I. Richter, A. Diedrich, G. Musmann, U. Auster, U. Motschmann, A. Balogh, C. Carr, E. Cupido, A. Coates, M. Rother, K. Swingenschuh, K. Szegö and B. Tsurutani, "RPC-MAG – The Fluxgate Magnetometer in the ROSETTA Plasma Consortium", *Space Science Reviews*, Vol. 128, pp. 649-670, 2007.
[22] M. Billvik, "The first Rosetta Earth flyby – Trajectory, attitude and radiation information for LAP operations", *Uppsala Universitet*, 2005.
[23] Y. Sugimoto, J. C. van der Ha and B. Rievers, "Thermal Radiation Model for the Rosetta Spacecraft", *AIAA 2010-7659*, 2010.
[24] U. De Filippis, C. Lefevre, M. Lucente, C. Magnafico and F. Santoli, "Characterization of the outgassing event during BepiColombo second Venus flyby using Italian Spring Accelerometer data", *Acta Astronautica*¸ 226, pp. 11-19, 2025.
[25] C. Magnafico, U. De Filippis, F. Santoli, C. Lefevre, M. Lucente, D. Lucchesi, E. Fiorenza, R. Peron and V. Iafolla, "Italian Spring Accelerometer measurements of unexpected Non Gravitational Perturbation during BepiColombo second Venus swing-by", *Acta Astronautica*, 232, pp. 14-22, 2025.